\documentclass[12pt,preprint]{aastex}
\usepackage{rotating}
\usepackage{multirow}
\usepackage{color}
\usepackage{CJK}

\newcommand{\tabincell}[2]{\begin{tabular}{@{}#1@{}}#2\end{tabular}}

\slugcomment{}
\shortauthors{Zhu et al.}
\begin{document}

\title{Empirical Study of Simulated Two-planet Microlensing Events}
\begin{CJK*}{UTF8}{gkai}

\author{Wei Zhu (祝伟) \altaffilmark{1,2,3}, Andrew Gould\altaffilmark{1}, Matthew Penny\altaffilmark{1,4}, Shude Mao\altaffilmark{2,4}, and Rieul Gendron\altaffilmark{4}}
%\author{Wei Zhu \altaffilmark{1,2,3}, Andrew Gould\altaffilmark{1}, Matthew Penny\altaffilmark{1,4}, Shude Mao\altaffilmark{2,4}, and Rieul Gendron\altaffilmark{4}}

\email{weizhu@astronomy.ohio-state.edu}

\altaffiltext{1}{Department of Astronomy, The Ohio State University, 140 W. 18th Ave., Columbus, OH 43210, USA}
\altaffiltext{2}{National Astronomical Observatories, Chinese Academy of Sciences, 20A Datun Road, Chaoyang District, Beijing 100012, China}
\altaffiltext{3}{Department of Astronomy, Peking University, Yi He Yuan Lu 5, Hai Dian District, Beijing 100871, China}
\altaffiltext{4}{Jodrell Bank Centre for Astrophysics, University of Manchester, Alan Turing Building, Manchester M13 9PL, UK}

\begin{abstract}
    We undertake the first study of two-planet microlensing models recovered from simulations of microlensing events generated by realistic multi-planet systems in which 292 planetary events including 16 two-planet events were detected from 6690 simulated light curves.  We find that when two planets are recovered, their parameters are usually close to those of the two planets in the system most responsible for the perturbations. However, in one of the 16 examples, the apparent mass of both detected planets was more than doubled by the unmodeled influence of a third, massive planet. This fraction is larger than, but statistically consistent with, the roughly 1.5\% rate of serious mass errors due to unmodeled planetary companions for the 274 cases from the same simulation in which a single planet is recovered.  We conjecture that an analogous effect due to unmodeled stellar companions may occur more frequently. For seven out of 23 cases in which two planets in the system would have been detected separately, only one planet was recovered because the perturbations due to the two planets had similar forms.  This is a small fraction (7/274) of all recovered single-planet models, but almost a third of all events that might plausibly have led to two-planet models.  Still, in these cases, the recovered planet tends to have parameters similar to one of the two real planets most responsible for the anomaly.
\end{abstract}


%%%%%%%%%%%%%%%%%%%%%%%%%%%%
\section{Introduction}

Microlensing is an important method in detecting extrasolar planets. It is sensitive to planets around the Einstein radius \citep{Mao:1991,Gould:1992}
\begin{equation}
\theta_{\rm E} \equiv \sqrt{\kappa M_{\rm L} \pi_{\rm rel}};\ \kappa \equiv \frac{4G}{c^2 \rm AU} \approx 8.14 \frac{\rm mas}{M_\odot}
\end{equation}
where $M_{\rm L}$ is the lens mass, $\pi_{\rm rel}={\rm AU} (D_{\rm L}^{-1}-D_{\rm S}^{-1})$ is the lens-source relative parallax, and $D_{\rm L}$ and $D_{\rm S}$ are the distances to the lens and source, respectively. For typical microlensing events, this angular scale corresponds to a physical size on the lens plane ($R_{\rm E}=D_{\rm L} \theta_{\rm E}$) that typically lies somewhat beyond the ``snow line'', the region that is believed to be the birth place of giant planets \citep[e.g.,][]{ida2004a}. To date, dozens of microlensing planets have been detected, enabling statistical studies of the frequency of distant bound planets \citep{Gould:2010,Sumi:2010,Cassan:2012}, and unbound ``free-floating planets'' \citep{Sumi:2011}.

However, among dozens of microlensing events that have yielded planet detections, only two systems securely contain two planets \citep{Gaudi:2008,Han:2013}, although most planets are believed to have been formed in multiple-planet systems. In addition, OGLE-2007-BLG-349 is either a two-planet event or a single planet system orbited by a stellar binary companion \citep{Dong:2014}. This low yield of multiple-planet systems via microlensing is the result of both the lower probability of detecting these events and the challenge in interpreting the collected data. The frequency of a microlensing event is $\sim 10^{-5}$ yr$^{-1}$ star$^{-1}$, of which only a few percent yield a planet detection. For example, the Optical Gravitational Lensing Experiment \citep[OGLE,][]{Udalski:2003} is monitoring $\sim 10^{8}$ stars in the Galactic Bulge, resulting in $\sim 2000$ microlensing events per year but only $\sim 10$ planet detections per year. The probability of detecting two-planet events is of order the product of the probability of detecting each of them \citep{Gaudi:1998,Zhu:2014}, meaning that the expected number of two-planet events based on the past 15-year observations is roughly a few, consistent with the number of these events that have been observed.

The interpretation of an observed two-planet event is challenging. For example, the first two-planet event OGLE-2006-BLG-109 \citep{Gaudi:2008,Bennett:2010} shows five distinct caustic-crossing features caused by a Jupiter/Saturn analog, with both finite source effects and orbital motion effects involved. Only with a thorough understanding about the caustic structures can one figure out that the first, second, third and fifth features are due to one planet while the fourth feature is due to the other \citep[see the Figure 2 of][]{Gaudi:2008}. In fact, unlike both OGLE-2006-BLG-109 and OGLE-2012-BLG-0026 for which the two planets both show noticeable and separated perturbations, two-planet events can show noticeable perturbation(s) of only one planet \citep[e.g., OGLE-2007-BLG-349,][]{Dong:2014}, making it even more difficult to find the correct two-planet (or triple-lens) model.

On the other hand, in the era of future microlensing experiments, the detection rate of multiple-planet events will increase significantly and thus the current understanding of such events must be improved in order to process the data in a timely way. Numerical simulations of the second-generation microlensing surveys [i.e., OGLE IV, the Microlensing Observations in Astrophysics II \citep[MOA,][]{Bond:2001}, and especially the Korea Microlensing Telescopes Network \citep[KMTNet,][]{Kim:2010}] suggest that the planet detection rate will then be increased by at least a factor of several \citep[e.g.,][]{Shvartzvald:2012,Zhu:2014,Henderson:2014}. Space-based microlensing surveys like the Wide Field InfraRed Space Telescope \citep[WFIRST,][]{Spergel:2013} and EUCLID \citep{Penny:2013} will be able to increase the rate by another factor of $\sim 10$. Studies of multiple-planet microlensing events are therefore a necessity.

Various studies have been conducted related to multiple-planet events. \citet{Gaudi:1998} pointed out that multiple-planet events are more likely to be found in high-magnification ($A_{\rm max}>100$) events although the joint perturbation pattern around the peak of the light curve makes them difficult to interpret. \citet{Han:2005} noticed that in many cases the central planetary perturbations induced by multiple planets can be well approximated by the superposition of the single-planet perturbations. \citet{Song:2014} studied the degeneracies in the triple-lens case from a theoretical aspect. Both \citet{Gaudi:1998} and \citet{Song:2014} found that the light curves produced by multiple-planet systems can sometimes be degenerate with that from a single planet system. This double/triple lens degeneracy may lead to wrong planetary parameters as well as underestimation of the frequency of multiple-planet systems \citep{Song:2014}.

The present work is a follow-up paper to \citet{Zhu:2014} (hereafter Paper I) in which microlensing simulations were conducted for realistic planetary systems drawn from planet population synthesis simulations. This paper is focused on the (candidate) multiple-planet events produced by our simulation. In \S2 we give an overview of our simulation. Results concerning the detectability of the second planet in the presence of the first planet, and the recovery of planetary parameters, are presented in \S3. We discuss our results in \S4.

%%%%%%%%%%%%%%%%%%%%%%%%%%%%
\section{Simulation Overview} \label{sec:overview}

In Paper I, we simulated 6690 microlensing events. The lens systems are drawn from the \citet{ida2004a,ida2004b,ida2005,ida2008a,ida2008b,ida2010} core accretion model for $0.3M_\odot$ stars. We extracted 669 systems with at least one planet more massive than $0.1M_\oplus$ from 1000 Ida \& Lin systems, based on which 6690 microlensing light curves for typical bulge microlensing events were simulated (10 for each system). Light curves were generated by inverse ray shooting \citep{Schneider:1986,Schneider:1987} and synthetic photometry generated to match the cadence and precision of KMTNet using the synthetic photometry code of \citet{Penny:2011} and blending distributions of \citet{Smith:2007}. See Paper I for more details.

Simulated light curves were first fitted with a standard single-lens model. We use $\Delta \chi^2_{\rm single}$ to denote the deviation in $\chi^2$ of the best-fit single-lens model from the simulated data. Events with $\Delta \chi^2_{\rm single}>200$ were picked out to generate single-planet (double-lens) light curves for the six most probably detectable planets. These single-planet light curves were also subjected to single-lens modeling to determine the detectability of each planet considered by itself, which we denote as $\Delta \chi^2_{\rm sp,i}$ with $i$ the ranking of the planet in that system. After the removal of ``bad'' events, for which the sources were so bright that the numerical precision of the ray shooting procedure was not adequate, and defining a planet as detectable when its $\Delta \chi^2_{\rm sp,i}>200$, we finally confirmed $292$ planetary events, of which $23$ have two detectable planets and none has more than two.
\footnote{In fact we have one planetary event with $\Delta \chi^2_{\rm sp,2},\ \Delta \chi^2_{\rm sp,3}\approx150$, but it cannot be claimed as a three-planet event with sufficient confidence.}

Double-lens modeling was performed for the $23$ two-planet event candidates to examine the double/triple lens degeneracy. The result of this double-lens modeling is denoted as $\Delta \chi^2_{\rm double}$, meaning the $\chi^2$ difference between the best-fit single-planet (double-lens) model and the theoretical model. With a threshold of $\Delta \chi^2_{\rm double}=300$ \footnote{See Paper I for the reason of choosing this threshold value.}, we then confirmed $16$ two-planet events out of the 23 candidate events; that is, the remaining $7$ events with two detectable planets suffers the double/triple lens degeneracy. Statistical studies and implications based on these planetary events and two-planet events were presented in Paper I.

%%%%%%%%%%%%%%%%%%%%%%%%%%%%
\section{Results}
\subsection{Detectability of the second planet}

We first investigate the influence of the first planet (primary perturber) on the detectability of the second planet (secondary perturber). We quantify the detectability of the second planet in the presence of the first planet as the difference in $\chi^2$ between the best-fit single-planet (double-lens) model and the input model, i.e., $\Delta \chi^2_{\rm double}$. As is described in Section \S\ref{sec:overview}, the detectability of the second planet when it is alone can be quantified by $\Delta \chi^2_{\rm sp,2}$, which is the $\Delta \chi^2$ between the single-planet light curve (produced by removing all the other planets except for planet 2) and its best-fit single-lens model. If $\Delta \chi^2_{\rm double} < \Delta \chi^2_{\rm sp,2}$, we consider the detectability of the second planet is suppressed due to the presence of the first planet. Otherwise it is enhanced.

\begin{figure}
\centering
\epsscale{1.0}
\plottwo{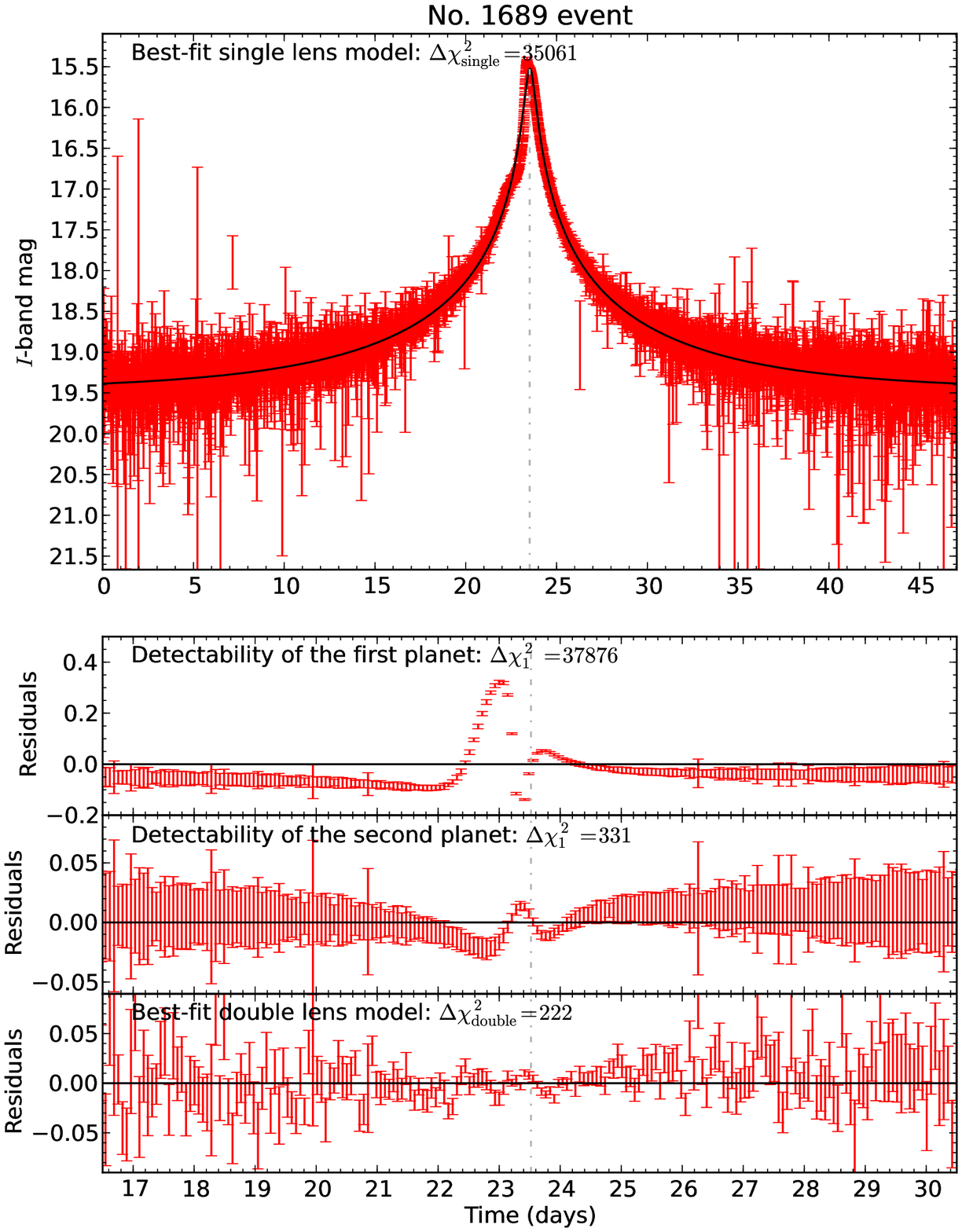}{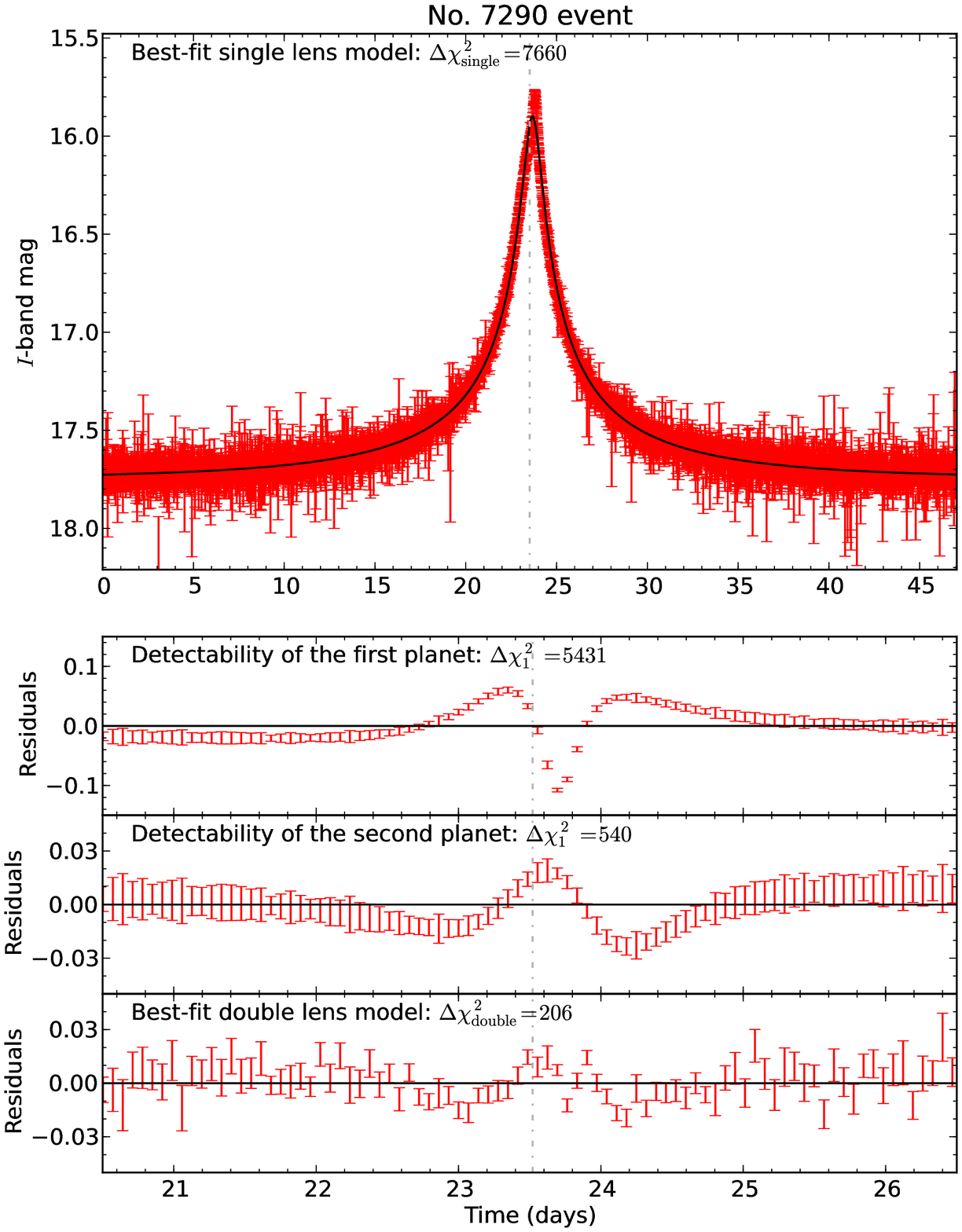}
\epsscale{1.0}
\plottwo{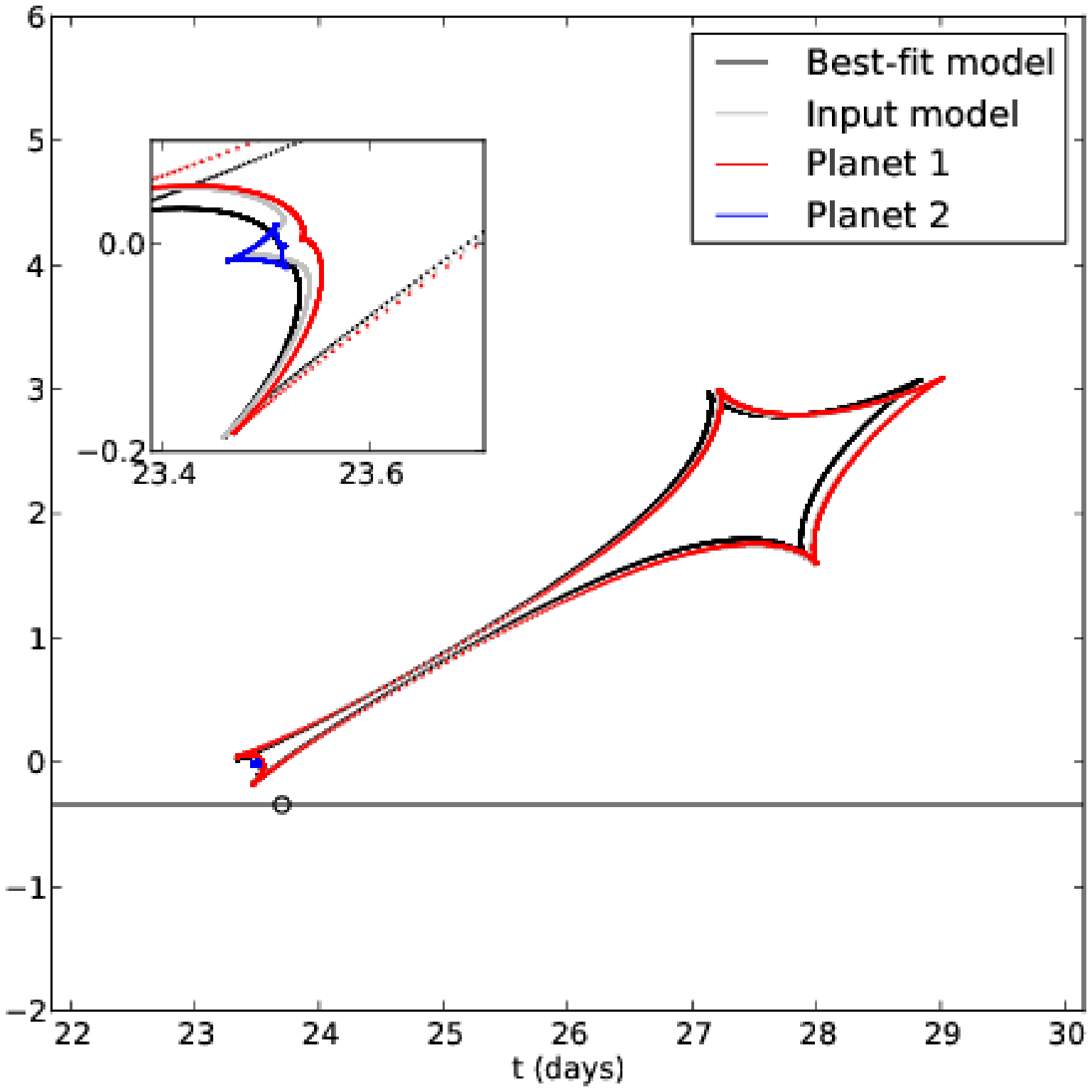}{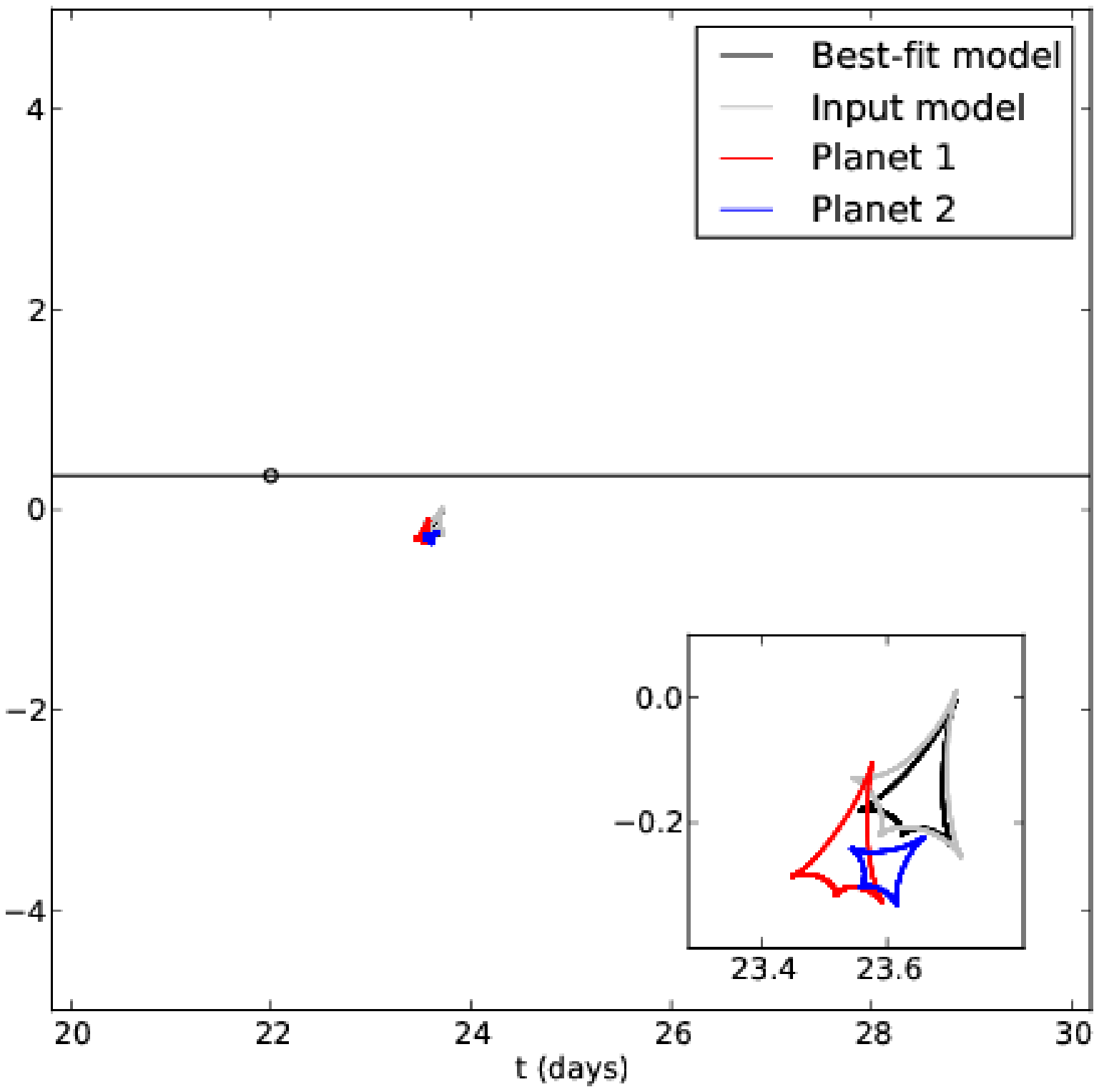}
\caption{Two examples of the double/triple lens degenerate events. For each event, we show from top to bottom the simulated data with the best-fit single-lens model, residuals between single-planet (planet 1) light curve and its best-fit single-lens model, residuals between single-planet (planet 2) light curve and its best-fit single-lens model, residuals between simulated data and its best-fit double-lens model, and caustic structures. Data points in residual plots are binned for clarity. Different colors in the caustic plots are: best-fit double-lens model (black), input triple-lens model (grey), caustics from star and only one planet (red and blue).
\label{fig:example-degenerate}}
\end{figure}

We relate the change in the detectability of the second planet to the similarity between the planetary perturbations caused by the two responsible planets when they are considered separately. We find that two-planet event candidates can be classified according to the similarity in perturbations. Events in which the two planetary perturbations appear at the same time (difference of less than 0.2 day) on the light curve and show either positive or negative correlations, are considered as Type I events. Otherwise they are considered as Type II events. We find that the detectability of the second planet is always suppressed by the presence of the first planet for Type I events, as is shown in Table~\ref{tab:chi2-all}. In fact, in $7$ of Type I events the suppression is so severe that the second planet is not a secure detection and therefore suffers the double/triple lens degeneracy. We show two of them in Figure~\ref{fig:example-degenerate}. For Type II events having noticeable planetary features induced by both planets, the detectability of the second planet is more or less conserved compared to the case when this planet is considered alone, suggesting that the detection of the two planets are independent. The published two-planet events, OGLE-2006-BLG-109 \citep{Gaudi:2008} and OGLE-2012-BLG-0026 \citep{Han:2013}, belong to this class, and an example of such event from our simulation is shown in Figure~3 of Paper I. For other Type II events, things are more complicated: the detectability of the second planet can either be suppressed or enhanced, and this cannot be distinguished on the basis of the caustic types by which the planets are detected. For example, the two planets in events No.4247 and No.8451 are detected by the resonant caustic (for planet 1) and the central caustic (for planet 2), but the detectability of the second planet is enhanced in No.4247 event while it is suppressed in No.8451 event.

With the results of double-lens modeling of 23 two-planet event candidates, we find that in most cases the residuals in data after the subtraction of the best-fit double-lens model show similar patterns with that caused by the second planet when it was alone, suggesting that the overall planetary perturbation is a superposition of perturbations caused by the two responsible planets. A similar conclusion has been reached by \citet{Han:2005}.

%%%%%%%%%%%%%%%%%%%%
\subsection{Parameter recovery}

The lens system consists of detectable and undetectable planets. The detectable planet(s) can distort the light curve of a single lens (the host star), while the undetectable planets can also affect the single-lens light curve by changing the center of magnification and the center of mass. Furthermore, due to the degeneracy existing between light curves arising from multiple- and single-planet events \citep{Gaudi:1998,Song:2014}, multiple detectable planets may not all be recovered from the light curve. In fact, as we have seen in Table~\ref{tab:chi2-all}, 7 events having two planets detectable individually suffer such double/triple lens degeneracy and thus the recovered planetary parameters might not be physical, although in principle they should be close to the true parameters of one of the detectable planets (the primary perturber). Here we investigate the deviation of the recovered planetary parameters from the true parameters.

\subsection{Single planet events}

\begin{figure}
\centering
\epsscale{0.8}
\plotone{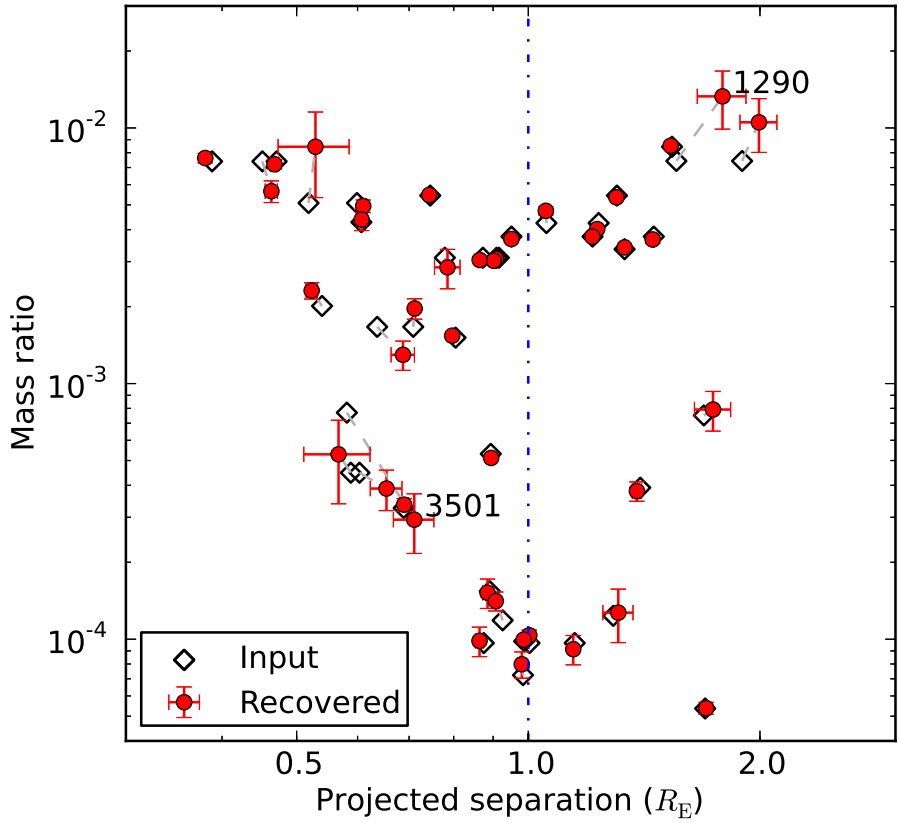}
\caption{The mass ratio $q$ and projected separation $s$ of the recovered planets (red filled circles) from the double-lens modeling of the light curves of 41 single-planet events, with respect to the true values of the detectable planets (open diamonds) in these events. Grey dashed lines connect the recovered planet with the original responsible planet.
\label{fig:recovery-single}}
\end{figure}

In Paper I we have mentioned that $85\%$ of all single-planet events can be reproduced very well by the corresponding single-planet light curve without any fitting. Here we investigate how the planetary parameters are perturbed in the remaining $15\%$ (i.e., 41) single-planet events. When we look into the planetary systems that cause these events, massive ($q>0.001$) and either far ($s>3$) away from or close ($s<0.3$) to the host star planets (in some cases brown dwarfs) appear in all these 41 cases. In addition, most of these events are non-caustic-crossing events. For all these 41 events, we fit a double-lens model to the simulated data, with the true planetary parameters as the initial seed. The best-fit planetary parameters, i.e., planet-to-star mass ratio $q$ and projected separation $s$, given by these double-lens models, are displayed in Figure~\ref{fig:recovery-single}, with the true parameters shown for comparison.

\begin{figure}
\centering
\epsscale{1.0}
\plotone{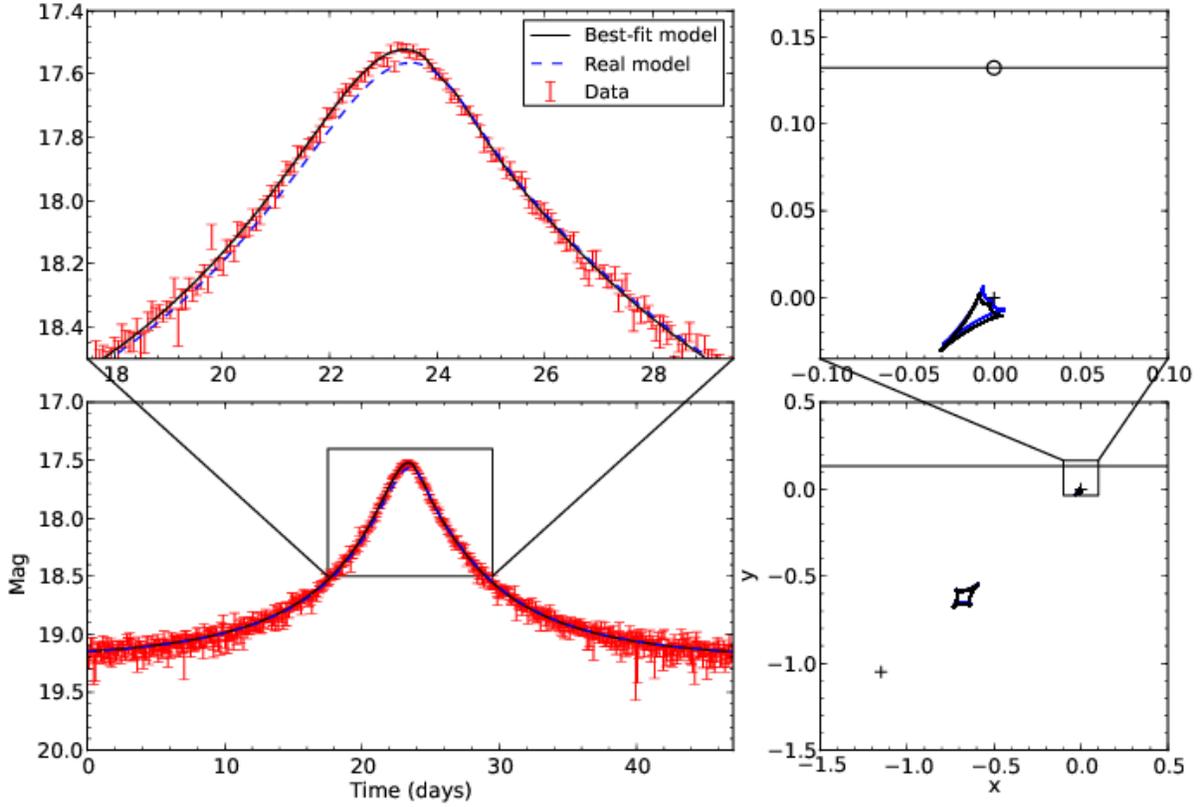}
\caption{The light curve and caustic structure of No.1290 event. In the light curve plot, the black solid line shows the best-fit double-lens light curve, while the blue dashed line is the light curve generated from the real planet. In the caustic structure plot, the black horizontal line is the source trajectory, with the circle centered on it as the size of the source, the plus signs indicate the positions of lenses, the blue curve is the caustic from the real planet, and the black curve is that from the real planet and an undetectable, distant ($s=14.6$) but massive ($q=0.14$) brown dwarf.
\label{fig:no-1290}}
\end{figure}

\begin{figure}
\centering
\epsscale{1.2}
\plotone{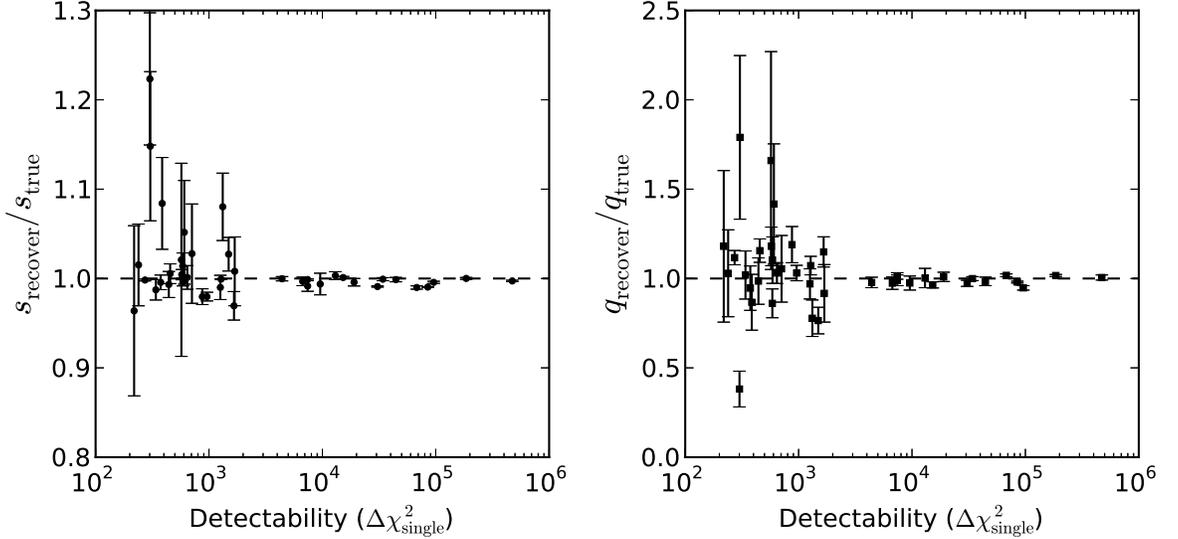}
\caption{Ratio between the recovered parameter and the true parameter vs. the detectability of the planet quantified by $\Delta \chi^2_{\rm single}$, with the projected separation $s$ shown on the left and the mass ratio $q$ on the right. The error bars are given by the Markov chain Monte Carlo (MCMC) modeling of the light curves. Note that it is the absolute deviation that is of interest, even though that for some planets with high $\Delta \chi^2_{\rm single}$ the deviation in parameters is statistically significant.
\label{fig:comparison-single}}
\end{figure}

Figure~\ref{fig:recovery-single} shows that even for events for which undetectable planets perturb the single-planet light curves, the planetary parameters recovered by modeling the light curve still represent the true parameters very well, although in very few cases, the planetary mass ratio can be off by a factor of two (e.g., No. 1290 and 3501 events, as are indicated in Figure~\ref{fig:recovery-single}), an example of which is shown in Figure~\ref{fig:no-1290}. These marginally detected planets appear in non-caustic-crossing events, and are therefore more easily affected by non-detectable planets in the same system, as is confirmed in Figure~\ref{fig:comparison-single} which shows the ratio between the recovered and true planetary parameters with respect to the detectability of the planet quantified by $\Delta \chi^2_{\rm single}$. On the other hand, planets in events with high $\Delta \chi^2_{\rm single}$ can all be recovered very well, since the very distinct structures on the light curves due to approaching or crossing the caustics are robust to perturbation by distant planets.

\subsubsection{Double/Triple lens degenerate events}

\begin{figure}
\epsscale{0.8}
\plotone{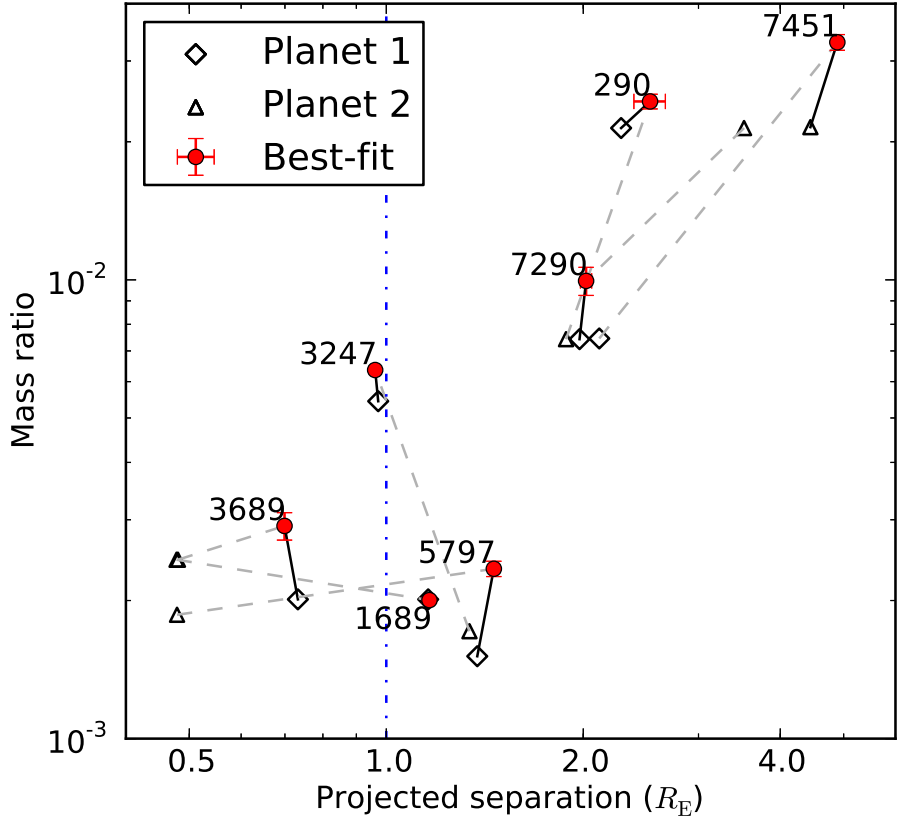}
\caption{The mass ratio $q$ and projected separation $s$ of the recovered planet (red filled circles) from the double-lens modeling of the light curves of 7 double/triple lens degenerate events, with respect to the true values of the two detectable planets (open diamonds and open triangles) in these events. Solid lines are used to connect the recovered planet with the closer one between the two detectable planets, and grey dashed lines connect it with the other one in the same event. The vertical dash-dotted line indicates the position of the Einstein ring radius.
\label{fig:recovery-degenerate}}
\end{figure}

Due to the similarity between the planetary perturbation patterns caused by the two detectable planets when considered separately, 7 out of the 23 two-planet event candidates do not have large enough $\Delta \chi^2_{\rm double}$ to claim the detection of the second planet. The planetary parameters that are recovered should therefore deviate from the values of either planet 1 or planet 2. We quote the closer one between the two detectable planets in the mass ratio-separation parameter space as the true position of the recovered planet. We list in Table~\ref{tab:degeneracy} and illustrate in Figure~\ref{fig:recovery-degenerate} the most important parameters, i.e., mass ratio $q$ and projected separation $s$, of the two detectable planets and the recovered planet. We find that the deviation in the projected separation ($s$) is below $\sim 10\%$, while the deviation in mass ratio ($q$) is within $35\%$, from the true planet. Therefore, even in events suffering the double/triple lens degeneracy, the planet recovered from modeling the microlensing light curve is still a good representative of one of the true planets.

\begin{figure}
\centering
\epsscale{0.8}
\plotone{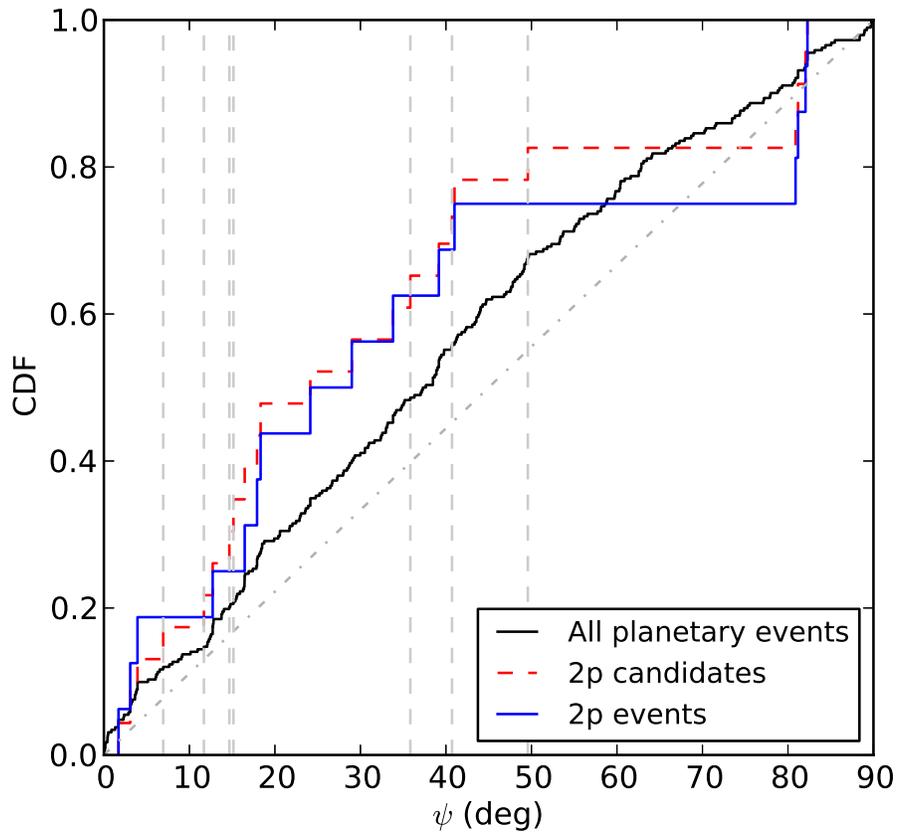}
\caption{The cumulative distributions of intersection angles between two most detectable planets in all planetary events (black solid line), between the two responsible planets in two-planet event candidates (red dashed line), and between two detected planets in confirmed two-planet events (blue solid line); vertical grey dashed lines indicate the positions of the seven double/triple lens degenerate events.
\label{fig:psi-dist}}
\end{figure}

We show in Figure~\ref{fig:psi-dist} the distributions of intersection angles between two responsible planets for two-planet events with (i.e., two-planet event candidates) and without (i.e., confirmed two-planet events) these double/triple lens degeneracy, with respect to the distribution of all planetary events, in which for single-planet events the intersection angle between two most detectable planets is used. A Kolmogorov-Smirnov test between the all planetary events sample and the two-planet event candidates sample shows that the probability of the two distribution being extracted from the same distribution is $22\%$, while it is $61\%$ between all planetary events sample and the confirmed two-planet events sample. The discrepancy between the results of these two KS tests suggests that there may be some preferred intersections angles for the double/triple lens degeneracy.

\subsubsection{Two-planet events}

\begin{figure}
\centering
\epsscale{0.8}
\plotone{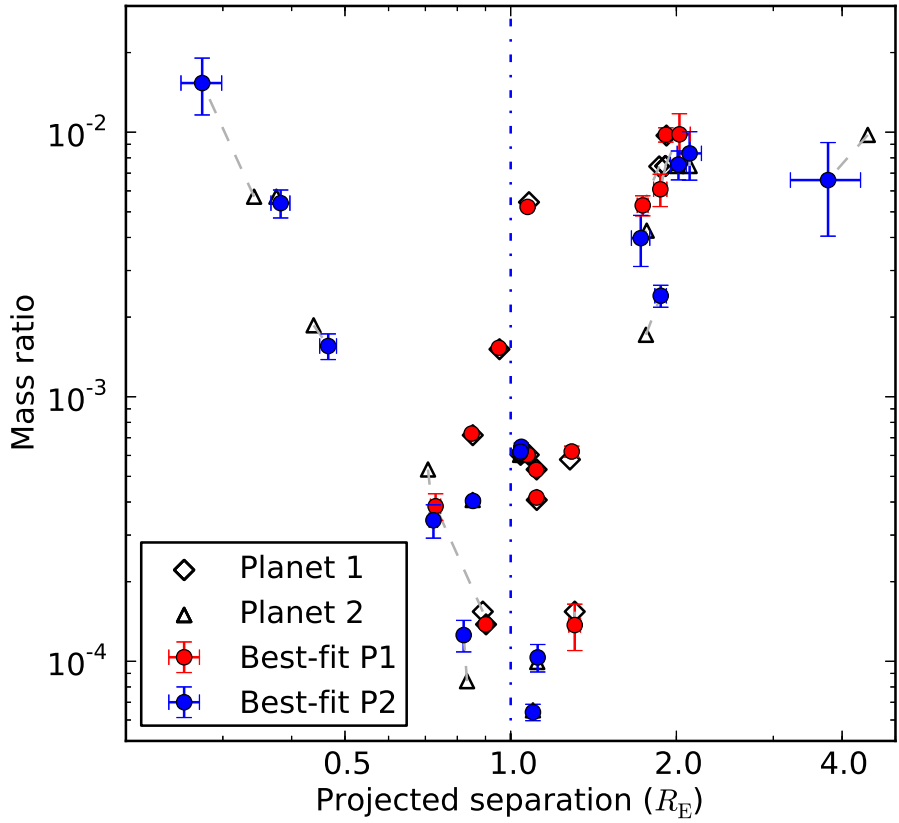}
\caption{The mass ratio $q$ and projected separation $s$ of the recovered planets (red and blue filled circles) from the triple-lens modeling of the light curves of 16 two-planet events, with respect to the true values of the two detectable planets (open diamonds and open triangles) in these events. Grey dashed lines connect the recovered planet and the original responsible planet.
\label{fig:recovery-2p}}
\end{figure}

\begin{figure}
\centering
\epsscale{1.2}
\plotone{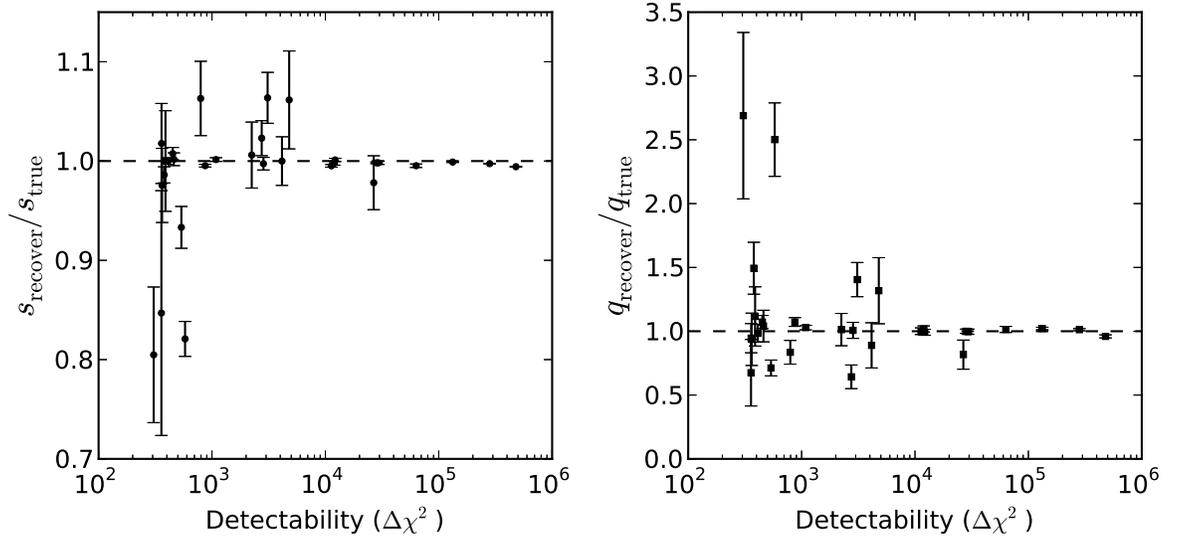}
\caption{Similar to Figure~\ref{fig:comparison-single} but for planets detected in two-planet events. For each event, the detectability of the first planet (the major perturber) is quantified by $\Delta \chi^2_{\rm single}-\Delta \chi^2_{\rm double}$, while that of the second planet (the minor perturber) is quantified by $\Delta \chi^2_{\rm double}$.
\label{fig:comparison-2p}}
\end{figure}

We also investigate the influence of the undetectable planets on the parameter recovery in two-planet events, by employing triple-lens fits to the simulated light curves. The recovered planetary parameters with respect to the original parameters for 16 two-planet events are shown in Figure~\ref{fig:recovery-2p}, and the ratio between these two as a function of the detectability of the planet is shown in Figure~\ref{fig:comparison-2p}.

\begin{figure}
\centering
\epsscale{1}
\plotone{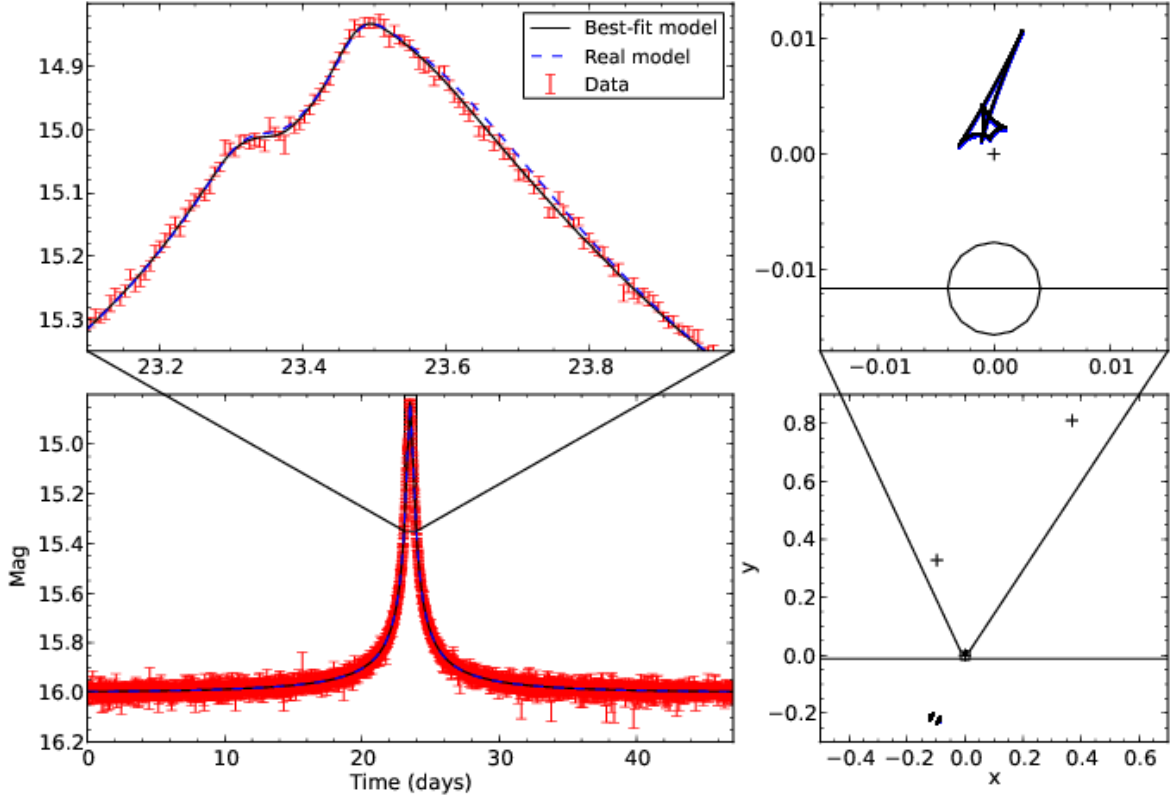}
\caption{Similar to Figure~\ref{fig:no-1290} but for No.8770 event. In this case, there are two detected planets, of which the second one is marginally detected ($\Delta \chi^2_{\rm double}=306$), and the distant massive perturber has $s=13.9$ and $q=0.003$.
\label{fig:no-8770}}
\end{figure}

Similar to what has been seen in the case of single-planet events, the two planets are recovered very well in most events, meaning that the remaining undetectable planets in these systems have a negligible influence on the two detectable planets, although the recovered planet masses in No.8770 event deviate from the true values by a factor of three, which also comes from the perturbation of a distant ($s\approx14$) massive ($q\approx3\times10^{-3}$) planet. The light curve and caustic structure of this event is shown in Figure~\ref{fig:no-8770}. As with Figure~\ref{fig:comparison-single} for the single-planet case, Figure~\ref{fig:comparison-2p} also indicates that the planets detected with lower $\Delta \chi^2$ in two-planet events are more easily affected by the undetectable planets in the same system.

The planetary close/wide degeneracy \citep{Dominik:1999,An:2005} has been found in two-planet events, both observationally \citep{Han:2013} and theoretically \citep{Song:2014}, in the case when two planets are both detected via central caustics. We find that the second planet in No.3194 event also suffers this close/wide degeneracy ($\Delta \chi^2=2$), despite the fact that the first planet has noticeable signatures caused by crossing its planetary caustic.

%%%%%%%%%%%%%%%%%%%%%%%%%%%%%
\section{Discussion}

In this work, by simulating microlensing light curves of realistic multi-planet systems, we are trying to answer two questions: 1) when will the detectability of a second planet be suppressed or enhanced in the presence of the first planet, and 2) whether and how can the other detectable or undetectable planets in the same lens system affect the parameter recovery of the detected planets.

We find that the influence of the first planet on the detectability of the second planet is strongly correlated with the similarity between the planetary perturbations caused by the two detectable planets when they are considered separately. Particularly, when the two detectable planets show similar perturbation patterns, the detectability of the second planet is always suppressed by the presence of the first planet; in some cases, such suppressions may be so strong that no signs of the perturbation caused by the second planet can be seen in the residuals after the subtraction of the best-fit double-lens (single-planet) model, meaning that these events suffer the double/triple lens degeneracy. Our result also confirms the well established result that the overall perturbation pattern in multiple-planet events is usually a superposition of the that of individual planets \citep{Han:2005}, indicating that the modeling of two-planet events can mostly be decomposed.

We compared the planetary parameters recovered from modeling the microlensing light curves with the input parameters, for single-planet events, double/triple lens degenerate events and double-planet events. In a large majority of the investigated events, the planetary parameters can be recovered very well. However, we also notice that in a very few low-magnification events in which the planets are marginally detected, the recovered planetary mass ratio can deviate from the true value by a factor of two to three, due to the existence of a distant massive companion to the lens star. The fraction of such events in the two-planet case (1/16) is larger than, but still statistically consistent with, that in the single-planet case (4/274), suggesting that the deviation in planetary parameters due to unmodeled massive planets does not become worse in the multiple-planet case. However, although such events contribute a very small fraction to the total planetary events in our simulation, we conjecture that this problem may be more serious for the case of more massive companions, namely stars.

In Paper I, we confirmed that follow-up observations are still necessary in extracting the full power of high-magnification events in the era of the second-generation microlensing experiments. In fact, follow-up observations are also necessary for the detection of multiple-planet systems. In two-planet events, for example, although the detectability of the second planet is suppressed by the presence of the first planet in most cases, this suppression can be compensated by the intensive follow-up observations which are more likely initiated by detecting the first planet, since the planetary perturbation of the first planet, the primary perturber, is larger and sustained longer. The much more intensive ($\sim 1$ min compared to 10 min of KMTNet) observations conducted by the follow-up teams \citep[e.g., $\mu$FUN,][]{Gould:2006,Gaudi:2008} can therefore compensate the suppression due to the first planet. Furthermore, with follow-up observations, one would expect to detect more multiple-planet events and more lower-mass planets than under the survey-only mode (Gould et al. 2014).

\acknowledgments
We would like to thank Shigeru Ida and Doug Lin for providing us the data from their population synthesis models, and Thijs Kouwenhoven and Rainer Spurzem for providing us their computing facilities. A.G. thanks the Chinese Academy of Sciences for a visiting professorship. This work has been partly supported by the Strategic Priority Research Program “The Emergence of Cosmological Structures” of the Chinese Academy of Sciences Grant No. XDB09000000, and by the National Natural Science Foundation of China (NSFC) under grant numbers 11333003 (SM and WZ). This work was also supported by NSF grant AST 1103471 (WZ and AG) and NASA NNGX12AB99G (AG).

%%%%%%%%%%%%%%%%%%%%%%%%%%

\clearpage
\begin{table*}
\centering
\caption{Model fitting results of the 23 two-planet event candidates, sorted by $\Delta \chi^2_{\rm double}$.
\label{tab:chi2-all}}
\small
\begin{tabular}{crrrllrcl}
\tableline\tableline
Classification & No. & $A_{\rm max}$ & $\Delta \chi^2_{\rm single}$ & \tabincell{c}{$\Delta \chi^2_{\rm sp,1}$\\caustic$^a$} & \tabincell{c}{$\Delta \chi^2_{\rm sp,2}$\\caustic$^a$} & $\Delta \chi^2_{\rm double}$ & $\Delta \chi^2_{\rm triple}$ & \tabincell{c}{Planet 2\\detectability} \\
\tableline
\multirow{12}{*}{\tabincell{c}{Type I\\(Similar\\patterns,\\12)}}
& 5797 & 25 & 5250 & 3497,c & 269 & 35 & -- & Suppressed$^*$\\
& 7451 & 19 & 477 & 1842,c & 355 & 117 & -- & Suppressed$^*$\\
& 3689 & 6 & 467 & 361,c & 330,c & 130 & -- & Suppressed$^*$\\
& 290 & 6 & 10395 & 4813,c & 1975,c & 161 & -- & Suppressed$^*$\\
& 3247 & 17 & 21509 & 18310,rc,+ & 1963 & 203 & -- & Suppressed$^*$\\
& 7290 & 35 & 7659 & 5430,c & 540 & 206 & -- & Suppressed$^*$\\
& 1689 & 55 & 35060 & 37876,rc,+ & 330 & 222 & -- & Suppressed$^*$\\
& 8770 & 74 & 885 & 824,c & 331,c & 306 & 2 & Suppressed\\
& 4225 & 122 & 63577 & 61675,c,+ & 675,c & 358 & 27 & Suppressed\\
& 4770 & 66 & 4502 & 3790,c & 742,c & 359 & 24 & Suppressed \\
& 3194 & 22 & 3206 & 4597,p+c,+ & 969,c & 363 & 19 & Suppressed\\
& 1797 & 126 & 282090 & 279535,rp+rc,+ & 821,c & 557 & 21 & Suppressed\\
\tableline
\multirow{11}{*}{\tabincell{c}{Type II\\(Different\\patterns,\\11)}}
& 5481 & 6 & 28655 & 31885,p & 255,c & 379 & 3 & Enhanced\\
& 6451 & 9 & 11701 & 9821,rp,+ & 765c & 389 & 11 & Suppressed\\
& 9262 & 11 & 135396 & 121313,r,+ & 792,c & 404 & 14 & Suppressed\\
& 8837 & 9 & 862 & 600,p & 371,c & 411 & 51 & --$^{**}$ \\
& 5290 & 24 & 998 & 1360,c & 548,c & 461 & 50 & Suppressed\\
& 3451 & 11 & 5667 & 4577,r,+ & 935,c & 872 & 19 & --$^{**}$ \\
& 8451 & 21 & 27776 & 28030,rc & 2301,c & 1085 & 33 & Suppressed\\
& 4247$^b$ & 61 & 482187 & 476646,rc,+ & 826,c & 1539 & 13 & Enhanced\\
& 9451 & 19 & 14182 & 11428,r & 2493,c & 2242 & 3 & --$^{**}$ \\
& 2247 & 304 & 187273 & 156704,rc,+ & 1873,c & 3897 & 38 & Enhanced\\
& 4262 & 23 & 41546 & 29974,r,+ & 10185,p & 12179 & 53 & --$^{**}$ \\
\tableline\tableline
\end{tabular}
\\ \scriptsize
\begin{description}
    \item[$^a$] ``Caustic'' here denotes the caustic type by which the planet is detected, with ``c'', ``p'',``rc'' and ``rp'' representing the central caustic, planetary caustic, central part and planetary part of the resonant caustic, respectively. Plus signs (``+'') indicate that this is a caustic-crossing event.
    \item[$^b$] No.4247 event is very similar to OGLE-BLG-349 \citep{Dong:2014}.
    \item[$^*$] These events are considered as double/triple degenerate events.
    \item[$^{**}$] These four events have noticeable and separate features caused by both planets.
\end{description}
\end{table*}

\begin{table*}
\centering
\caption{Perturbing planets and the best-fit single planet in the seven double/triple degenerate events; deviations in $q$ and $s$ is defined as the fractional error in $q$ and $s$ from the closest planet.
\label{tab:degeneracy}}
\small
\begin{tabular}{rrccccllc}
\tableline\tableline
No. & $A_{\rm max}$ & $q_1$ ($\times 10^{-3}$) & $s_1$ & $q_2$ ($\times 10^{-3}$) & $s_2$ & $q$ ($\times 10^{-3}$) & $s$ & \tabincell{c}{Deviations in $s$ and $q$\\ from the closest planet}\\
\tableline
290 & 7 & $21.4$ & $2.28$ & $7.43$ & $1.89$ & $23.9 \pm 0.9$ & $2.53 \pm 0.14$ & $9.7\%$, $10\%$\\
1689 & 56 & $2.01$ & $1.16$ & $2.45$ & $0.48$ & $2.00 \pm 0.03$ & $1.164 \pm 0.002$ & $0.4\%$, $0.6\%$\\
3247 & 18 & $5.44$ & $0.97$ & $1.71$ & $1.33$ & $6.32 \pm 0.05$ & $0.9616 \pm 0.0005$ & $1.1\%$, $14\%$\\
3689 & 7 & $2.01$ & $0.73$ & $2.45$ & $0.48$ & $2.9 \pm 0.2$ & $0.699 \pm 0.006$ & $4.8\%$, $31\%$\\
5797 & 25 & $1.51$ & $1.38$ & $1.86$ & $0.48$ & $2.34 \pm 0.09$ & $1.46 \pm 0.01$ & $5.7\%$, $35\%$\\
7290 & 36 & $7.43$ & $1.97$ & $21.4$ & $3.52$ & $9.85 \pm 0.7$ & $2.02 \pm 0.04$ & $2.3\%$, $25\%$\\
7451 & 19 & $7.45$ & $2.11$ & $21.5$ & $4.45$ & $31.9 \pm 1.3$ & $4.89 \pm 0.02$ & $9.0\%$, $33\%$\\
\tableline\tableline
\end{tabular}
\normalsize
\end{table*}

\end{CJK*}
\end{document}